\documentclass[a4paper,11pt]{article}
\pdfoutput=1 

\usepackage{jheppub} 
\usepackage{slashed}
\usepackage[T1]{fontenc} 
\usepackage{graphicx}
\usepackage{epsfig}
\usepackage{amsmath,amssymb}
\usepackage{MnSymbol}
\usepackage{physics}
\def\p{\partial}

\def\ee{\end{equation}}
\def\be {\begin{equation}}

\def\zb{{\bar z}}

\def\kap{\kappa}
\def\sig{\sigma}
\def\lam{\lambda}
\def\half{\frac{1}{2}}

\numberwithin{equation}{section}

\title{\boldmath Perturbatively  Exact $w_{1+\infty}$ Asymptotic Symmetry of Quantum Self-Dual Gravity }

\author[a]{Adam Ball,}
\author[a]{Sruthi A. Narayanan,}
\author[a,b]{Jakob Salzer}
\author[a]{and Andrew Strominger}

\affiliation[a]{Center for the Fundamental Laws of Nature, Harvard University,\\
Cambridge, MA 02138, USA} 
\affiliation[b]{Physique Th\'eorique et Math\'ematique, Universit\'e libre de Bruxelles and\\
International Solvay Institutes, Campus Plaine C.P. 231, B-1050
Bruxelles, Belgium
}

\abstract{The infinite tower of positive-helicity soft gravitons  in any minimally coupled, tree-level, asymptotically flat four-dimensional (4D) gravity was recently shown to generate a $w_{1+\infty}$ asymptotic symmetry algebra. It is natural to ask whether this classical algebra acquires  quantum corrections at loop level. We explore this in quantum self-dual gravity, whose amplitudes acquire known one-loop exact all-plus helicity quantum corrections. We show using collinear splitting formulae that, remarkably, the $w_{1+\infty}$ algebra persists in quantum self-dual gravity without corrections.}
\begin{document} 
\maketitle
\flushbottom

\section{Introduction}

Any 4D quantum theory of gravity in asymptotically flat (AF) spacetime admits the non-trivial action of an infinite number of asymptotic symmetries arising from soft theorems~\cite{Strominger:2013jfa,He:2014laa}. These symmetries organize into infinite towers  corresponding to terms in a soft expansion with supertranslations at the top, followed by superrotations at subleading order. Recently it was shown~\cite{Guevara:2021abz,Strominger:2021lvk} that at tree level, the entire tower of positive-helicity soft graviton symmetries in minimally coupled gravity is generated by a $w_{1+\infty}$ algebra\footnote{As explained below, more precisely it is  a $w_{1+\infty}$-wedge current algebra, but we refer to it as a $w_{1+\infty}$ algebra for brevity.}. 

One wishes to understand whether the $w_{1+\infty}$ algebra is exact,  deformed, or anomalous in the full theory with both non-minimal couplings and quantum corrections.  Possible classical or quantum deformations of the theory are severely restricted by various constraints. In low-energy effective field theory, there are only three operators which can classically correct any of the first three soft symmetries~\cite{Elvang:2016qvq} (whose commutators generate the whole tower). At the quantum level, further corrections may arise from loops of massless particles. 

In  2D systems with a $w_{1+\infty}$ symmetry, such as a free fermion or boson, quantum corrections were indeed found to deform the algebra. However, the Jacobi identity allows only a two parameter family of deformations, to an algebra known as $W_{1+\infty}$, see~\cite{Pope:1991ig,Shen:1992dd}. Hence a mathematically consistent outcome  is that the classical $w_{1+\infty}$ symmetry is deformed to $W_{1+\infty}$ in the full theory of 4D quantum gravity, but this does not occur in the example considered here. 

In this paper we consider a very simple  toy model in which quantum corrections to $w_{1+\infty}$ can be very explicitly studied: quantum self-dual gravity in $(2,2)$ signature Klein space~\cite{Ooguri:1991fp, Chalmers:1996rq,Bern:1998xc,Bern:1998sv,Krasnov:2016emc}. At tree level, this theory has been known to have a $w_{1+\infty}$ symmetry since early work of Penrose on twistors~\cite{Penrose:1968me,Penrose:1976js,Boyer:1985aj}. It has recently been shown~\cite{Adamo:2021lrv} that the generic soft gravitational $w_{1+\infty}$ symmetry of~\cite{Strominger:2021lvk} reduces to the one considered by Penrose when specialized to the twistor context, see also~\cite{Costello:2021bah}. At the quantum level, perturbative corrections to self-dual gravity are
one-loop exact and appear only in the all-plus helicity sector. The somewhat lengthy  explicit expression for the all-plus amplitude first appeared in~\cite{Bern:1998xc}. 

Our main result herein  is  that $w_{1+\infty}$ persists as an exact symmetry of the all-orders quantum self-dual gravity scattering amplitudes.

This is a proof of concept that quantum loops need not topple the intricate tower of soft symmetries possessed by any minimally coupled theory of gravity at tree level~\cite{Himwich:2021dau}. But it would be hasty to expect similar results for a generic theory of 4D AF quantum gravity. Due, in part, to the vanishing of the unitarity cuts, the all-plus amplitudes are rational functions of the momenta with only soft and collinear singularities. Indeed this was the basis for their original construction in~\cite{Bern:1998xc}.  Hence the self-dual case is very special, and does not include the more complex analytic behavior expected in the generic case. Our result is closely related to that of~\cite{He:2014bga,Bern:2014oka} who showed that the leading, subleading, and subsubleading energetically soft theorems do not receive corrections for the all-plus amplitudes.\footnote{The results of~\cite{He:2014bga} could likely be augmented to prove the persistence of  $w_{1+\infty}$, but it would require a careful translation between the energetically soft  and conformally soft expansions. For this reason the proof given below by direct computation of the splitting function seemed quicker.} See also \cite{Bern:2014oka} for further discussion of loop corrections to soft theorems.

The paper is organized as follows. In section~\ref{sec:conventions} we detail the conventions for the metric and spinor-helicity variables used throughout the paper. In section~\ref{sec:winfty} we review how the $w_{1+\infty}$ algebra arises in a conformal basis starting from the OPE of positive-helicity graviton operators. In section~\ref{sec:selfdual} we briefly review the basics of quantum self-dual gravity. In particular, we present  the $\mathcal{S}$-matrix elements which we use to show  the algebra is undeformed at loop level. In section~\ref{sec:collinear} we show that since the collinear splitting functions remain unchanged at loop level in self-dual gravity, the algebra is perturbatively exact.

\section{Conventions}\label{sec:conventions}

We will work in $(-,+,-,+)$ signature ``Klein space'' throughout. Momenta for massless particles are parametrized as
\begin{equation}
p^\mu = \epsilon \, \omega \hat{p}^\mu = \epsilon \, \omega(1+z\bar{z}, z+\bar{z}, z-\bar{z}, 1-z\bar{z}) 
\end{equation}
where $\epsilon=\pm 1$ for outgoing and incoming particles respectively. This differs from the usual conventions by a factor of $i$ in the third component since we are in Klein  rather than Minkowski space. In Klein space $z,\bar{z}$ are independent real coordinates rather than complex conjugates and cover a signature $(1,1)$ Minkowski space $\mathbb{M}^2$. The celestial torus at null infinity is comprised of two copies of $\mathbb{M}^2$, corresponding to $\epsilon= 1$ or $\epsilon= -1$, glued together at their boundaries. 
In Klein space the Lorentz group becomes $\textrm{SO}^+(2,2)= \frac{\textrm{SL}(2,\mathbb{R})_L\times \textrm{SL}(2,\mathbb{R})_R}{\mathbb{Z}_2}$,\footnote{This is the identity component of ${\rm SO}(2,2)$. } which acts as the conformal group on the celestial torus:
\begin{equation}
\begin{aligned} \textrm{SL}(2,\mathbb{R})_L:&~~~~z\to {az+b \over cz+d}, ~~~~\bar{z}\to \bar{z},~~~~ad-bc=1, \\  \textrm{SL}(2,\mathbb{R})_R:&~~~~\bar{z}\to {\bar{a}\bar{z}+\bar{b}\over \bar{c}\bar{z}+\bar{d}},~~~~ z \to z ,~~~~ \bar{a}\bar{d}-\bar{b}\bar{c}=1. 
\end{aligned}
\end{equation}
Discussions of the global structure of and scattering in Klein space can be found in~\cite{Atanasov:2021oyu,Mason:2005qu}.

A basis for the Kleinian Pauli matrices is 
\begin{equation}
\sigma_{\mu} = \left(\begin{bmatrix} 1 & 0\cr 0 & 1\end{bmatrix}, \begin{bmatrix} 0 & 1\cr 1 & 0\end{bmatrix}, \begin{bmatrix} 0 & 1\cr -1 & 0\end{bmatrix}, \begin{bmatrix} -1 & 0\cr 0 & 1\end{bmatrix}\right) 
\end{equation}
which satisfy
\begin{equation}
\sigma_{\alpha\dot{\alpha}}^\mu\sigma_{\beta\dot{\beta}}^\nu\varepsilon^{\alpha\beta}\varepsilon^{\dot{\alpha}\dot{\beta}} = -2\eta^{\mu\nu} 
\end{equation}
where $\varepsilon$ is the two-dimensional Levi-Civita tensor with $\varepsilon^{12} = \varepsilon^{\dot 1\dot 2} = +1$. Defining $(p\cdot\sigma)_{\alpha\dot{\alpha}} = p_{\alpha\dot{\alpha}} = \lambda_{\alpha}\tilde{\lambda}_{\dot{\alpha}}$ we make the standard choice for spinors
\begin{equation}\label{spinorsorig}
\lambda_\alpha = \epsilon\sqrt{2\omega}\begin{bmatrix} z\cr 1\end{bmatrix}, \ \ \ \ \tilde{\lambda}_{\dot{\alpha}} = \sqrt{2\omega}\begin{bmatrix} \bar{z}\cr 1\end{bmatrix}.
\end{equation}
The spinor-helicity variables in our conventions are 
\begin{equation}
\langle ij\rangle = \lambda_\alpha^i\varepsilon^{\alpha\beta}\lambda_\beta^j = 2\epsilon_i\epsilon_j\sqrt{\omega_i\omega_j}z_{ij}, \ \ [ij] = \tilde{\lambda}_{\dot{\alpha}}^i\varepsilon^{\dot{\alpha}\dot{\beta}}\tilde{\lambda}^j_{\dot{\beta}} = 2\sqrt{\omega_i\omega_j}\bar{z}_{ij} 
\end{equation}
where $z_{ij}=z_i-z_j$. These relations imply, in addition, that $\langle ij\rangle[ij]=-2p_i\cdot p_j$. 

\section{Review of $w_{1+\infty}$ in 4D Gravity}\label{sec:winfty}

The simplest realization of $w_{1+\infty}$ symmetry is in a conformal primary basis of particle states in 4D gravitational scattering amplitudes. Operators corresponding to such states can be constructed from Mellin transforms of massless momentum eigenstates. These operators are labelled by a conformal dimension $\Delta$ and a point $(z,\bar{z})$ on the celestial torus. Following the conventions in~\cite{Pate:2019lpp, Guevara:2021abz,Strominger:2021lvk,Himwich:2021dau} we let $G^+_\Delta(z,\bar{z})$ denote an outgoing positive-helicity conformal primary graviton operator. The tree-level celestial OPE of outgoing positive-helicity gravitons in any minimally coupled theory of gravity is
\be \label{eq:GOPE} 
G^+_{\Delta_1}(z_1, \zb_1) G^+_{\Delta_2}(z_2, \zb_2) \sim -\frac{\kap}{2z_{12}} \sum_{n=0}^\infty B(\Delta_1-1+n,\Delta_2-1) \frac{\zb_{12}^{n+1}}{n!} \bar\p^n G^+_{\Delta_1+\Delta_2}(z_2, \zb_2). 
\ee 
It can be derived from the collinear limit of two gravitons in a momentum-space amplitude~\cite{Fan:2019emx}. A family of conformally soft positive-helicity gravitons is defined by 
\begin{equation}\label{dcv}
H^k(z,\zb) = \lim_{\varepsilon\rightarrow 0}\varepsilon G_{k+\varepsilon}^+(z,\zb), \ \ k= 2, 1, 0, -1,\ldots\,. 
\end{equation}
The weights are 
\begin{equation}
\left(h,\bar{h}\right) = \left(\frac{k+2}{2}, \frac{k-2}{2}\right). 
\end{equation}
An insertion of a conformally soft operator $H^k$ into a correlation function of conformal primary operators yields the leading conformally soft graviton theorem for $k=1$, the subleading for $k=0$, etc. The $k=2$ operator is necessary for the closure of the algebra but $H^{k=2}$ lies in the ideal of the algebra and so can be, and usually is, consistently set to zero.  

Restricting both operators in~\eqref{eq:GOPE} to be conformally soft isolates the poles of the Euler beta function. This leads to the OPE of the soft operators $H^k(z,\zb)$~\cite{Guevara:2021abz},
\begin{equation}
    \label{eq:fullHOPE}
    H^{k}(z_1,\zb_1)H^{\ell}(z_2,\zb_2)\sim -\frac{\kappa}{2z_{12}}\sum^{1-k}_{n=0}\binom{2-k-\ell-n}{1-\ell}\frac{\zb_{12}^{n+1}}{n!}\bar{\partial}^nH^{k+\ell}(z_2,\zb_2). 
\end{equation}
The operators~\eqref{dcv} have a mode expansion which is truncated\footnote{The $\rm{SO}^+(2,2)$-invariant norm vanishes for all but finitely many modes of $H^k(z,\zb)$.} in the case of the $\bar h \le 0$ soft operators as
\begin{equation}
\label{eq:Hmodes}
H^k(z,\bar{z}) = \sum_{n=\frac{k-2}{2}}^{\frac{2-k}{2}}\frac{H_n^k(z)}{\bar{z}^{n+\frac{k-2}{2}}}, 
\end{equation}
such that for each $k\le 2$ the $H^k_n(z)$ transform in a $(3-k)$-dimensional representation of $\textrm{SL}(2,\mathbb{R})_R$. 
Conversely, we can extract the mode $H^{k}_n(z)$ by the contour integral
\begin{equation}
    \label{eq:Hcont}
    H^k_n(z)=\oint_{|z|<\varepsilon}\frac{d \bar{z}}{2\pi i} \bar{z}^{n+\frac{k-4}{2}}H^k(z,\bar{z}). 
\end{equation} 
Each of these $H_n^k(z)$ is a 2D symmetry-generating conserved current. Their OPE follows from~\eqref{eq:fullHOPE} and was derived in~\cite{Guevara:2021abz}. It reads
\begin{equation}\label{eq:HmOPE}
\begin{aligned}  H^k_m&(z_1) H^\ell_n(z_2) \sim\cr
&- \frac{\kap}{2} [n(2-k) - m(2-\ell)] \frac{\left(\frac{2-k}{2}-m+\frac{2-\ell}{2}-n-1\right)!}{\left(\frac{2-k}{2}-m\right)!\left(\frac{2-\ell}{2}-n\right)!}\frac{\left(\frac{2-k}{2}+m+\frac{2-\ell}{2}+n-1\right)!}{\left(\frac{2-k}{2}+m\right)!\left(\frac{2-\ell}{2}+n\right)!} \frac{H^{k+\ell}_{m+n}(z_2)}{z_{12}}.
\end{aligned}
\end{equation}
This drastically simplifies with the redefinition~\cite{Strominger:2021lvk}
\be \label{wdef} 
w^p_m(z) \equiv \frac{1}{\kap} (p-m-1)! (p+m-1)! H^{4-2p}_m(z). 
\ee
As shown in~\cite{Jiang:2021ovh, Himwich:2021dau}, this is equivalent to using the modes of the $\bar{z}$-light transform of $H^{4-2p}(z,\zb)$, up to a mode-independent normalization. This leads to the OPE
\be \label{eq:wOPE}
w^p_m(z_1) w^q_n(z_2) \sim \frac{[m(q-1) - n(p-1)] w^{p+q-2}_{m+n}(z_2)}{z_{12}}.\ee
The index ranges are now
\be 
1-p \le m \le p-1 \quad {\rm with} \quad p=1,\frac{3}{2},2,\frac{5}{2},\dots \,.  \
\ee
The OPE~\eqref{eq:wOPE} takes the form of a level zero current algebra. The coefficient of $\frac{w^{p+q-2}_{m+n}(z_2)}{z_{12}}$ is encoded in the zero mode algebra of the $w^p_m(z)$, or equivalently in the celestial  commutator defined by 
\be \label{eq:wcom} \begin{aligned} {} [w^p_m, w^q_n](z) & \equiv \oint_z \frac{dw}{2\pi i} w^p_m(w) w^q_n(z) \\
& = [m(q-1) - n(p-1)] w^{p+q-2}_{m+n}(z). \end{aligned} 
\ee
This can be recognized as the wedge subalgebra\footnote{A more familiar example of this that we see in 2D CFT is $\mathfrak{sl}(2,\mathbb{C})$ as the wedge subalgebra of Virasoro. The full $w_{1+\infty}$ algebra corresponds to removing the restriction $|m| \le p-1$, so that $m \in \mathbb{Z} + p$.} of $w_{1+\infty}$~\cite{Bakas:1989xu} (although here $p$ is half-integral, not integral). Precisely this algebra appeared, with a rather different type of derivation,  in  Penrose's original twistor construction~\cite{Penrose:1968me,Penrose:1976js,Boyer:1985aj}. In fact, as discussed below,  it has recently been shown~\cite{Adamo:2021lrv} that the generic soft gravitational $w_{1+\infty}$ symmetry of~\cite{Strominger:2021lvk} reduces to the one considered by Penrose when specialized to the twistor context.  In particular they showed that the self-dual metric perturbations parametrized by a polynomial basis of area-preserving diffeomorphisms are precisely the conformal primary wavefunction modes defining our $w^p_m(z)$ operators. The algebra has appeared in  other places including  $W_{1+\infty}$-gravity~\cite{Pope:1991ig,Shen:1992dd} and the $c=1$ string~\cite{Klebanov:1991hx}.

\section{Quantum Self-Dual Gravity}\label{sec:selfdual}

At the classical level, self-dual gravity is defined by requiring the Riemann tensor to be self-dual. In $(2,2)$ Klein space  this means
\be \label{eq:SDEOM} 
R_{\mu\nu\rho\sig} = \half {\varepsilon_{\mu\nu}}^{\alpha\beta} R_{\alpha\beta\rho\sig}. 
\ee 
The anti-self-dual condition differs by a sign; for brevity we consider only the self-dual case herein.  The Lorentzian version is related by a Wick rotation, which introduces a factor of $\pm i$. Consequently there are no real solutions in Lorentzian signature. Hence, Kleinian spacetimes  provide a natural home for self-dual gravity.\footnote{Of course Euclidean self-dual gravity is also of great interest, but as there are no propagating gravitons, it is not so useful for the study of amplitudes.} Self-duality  implies Ricci flatness, so self-dual spacetimes are a special case of vacuum spacetimes. The simplest nontrivial self-dual solution is a positive-helicity plane wave, which is a linearized solution in Cartesian coordinates and an exact solution in Brinkmann coordinates~\cite{Brinkmann:1925}. Negative-helicity plane waves are anti-self-dual and are not solutions of~\eqref{eq:SDEOM}. Intuitively one can think of a general self-dual solution as built up from positive-helicity plane waves~\cite{Porter:1983gi}.

In this paper we wish to consider quantum self-dual gravity in Klein space. Several different definitions/formulations of quantum self-dual gravity exist in the literature, e.g.~\cite{Ooguri:1991fp, Chalmers:1996rq,Bern:1998xc,Krasnov:2016emc}, all of which reproduce the classical equation of motion~\eqref{eq:SDEOM}. At loop level some suffer from inconsistencies or other problems~\cite{Chalmers:1996rq}. Here we adopt the definition of quantum  self-dual gravity scattering amplitudes used in~\cite{Chalmers:1996rq,Bern:1998xc}. The action contains a Lagrange multiplier imposing the self-duality constraint, and the Lagrange multiplier itself is physically interpreted as the negative-helicity graviton. It shows up in a limited but nontrivial capacity: amplitudes contain  one or zero negative-helicity legs and all other helicities must be positive.\footnote{Some authors adopt  the point of view that the $\cal S$-matrix consists only of the scattering of positive-helicity gravitons, in which case the tree-level amplitudes vanish and all-plus scattering occurs only at one loop. For our purposes it is of interest to also consider the tree-level terms with one minus insertion because it affords a direct comparison of tree- and loop-level soft relations.}  The single-minus amplitudes are tree exact, while the all-plus amplitudes are one-loop exact and vanish at tree level.\footnote{As far as we are aware, nonperturbative effects have not been ruled out. }$^,$\footnote{Since the Lagrange multiplier appears linearly, it can be regarded as the loop-counting parameter after an appropriate rescaling. This shows that an $L$-loop amplitude has $1-L$ negative-helicity gravitons and $L\le 1$.} The only nontrivial tree-level amplitude is the MHV three-point amplitude
\be
M^{\rm tree}_3(1^+, 2^+, 3^-) = -\frac{i\kap}{2} \left( \frac{[12]^3}{[23][31]} \right)^2
\ee
which is kinematically forbidden in Lorentzian signature. It can be  shown that this tree-level $\cal S$-matrix, despite having only two- and three-point (connected) scattering, generates the full nonlinear space of classical solutions, as in the self-dual gauge theory case~\cite{Ooguri:1991fp,Cangemi:1996rx}. The all-plus amplitudes are nonzero for $n \ge 4$ legs, and they are both UV and IR finite. All the amplitudes of this formulation of self-dual gravity coincide with the corresponding amplitudes of Einstein gravity, and in the context of Einstein gravity they are often referred to as the ``self-dual sector''~\cite{Chalmers:1996rq,Bern:1998xc}. 

The one-loop all-plus $n$-graviton amplitudes were first found in~\cite{Bern:1998xc}. For arbitrary $n$ they are explicitly\footnote{In fact, this form of the amplitude has been explicitly checked only up to $n\le 6$. However, the above ansatz passes a number of non-trivial checks~\cite{Bern:1998xc}.} 
\begin{equation}
M_n^{1-\rm loop}(1^+,2^+,\ldots,n^+) = -\frac{i}{(4\pi)^2\cdot 960}\left(-\frac{\kappa}{2}\right)^n\sum_{\substack{1\leq a<b\leq n \\ M, N}} h(a,M,b)h(b,N,a)\mbox{tr}^3[aMbN] 
\end{equation}
where $a$ and $b$ label external legs, $M$ and $N$ are sets such that $M\cap N=\emptyset$ and $M\cup N = \{1,\ldots a-1,a+1,\ldots,b-1,b+1,\ldots,n\}$, and the choice of $(M,N)$ is not distinct from $(N,M)$.  The trace is defined as 
\be
\mbox{tr}[aMbN] = \langle a|K_M|b]\langle b|K_N|a]+[a|K_M|b\rangle[b|K_N|a\rangle
\ee
where $K_M = \sum_{i\in M} k_i$. The ``half-soft'' function that appears in the amplitude is related to BGK tree amplitudes~\cite{Berends:1988zp} and is defined as 
\begin{equation}
h(a,\{1,2,\ldots,n\},b) = \frac{[12]}{\langle 12\rangle}\frac{\langle a|K_{1,2}|3] \langle a|K_{1,3}|4]\cdots\langle a|K_{1,n-1}|n]}{\langle 23\rangle\langle 34\rangle\cdots \langle n-1,n\rangle\langle a1\rangle\langle a2\rangle\langle a3\rangle\cdots\langle an\rangle\langle 1b\rangle\langle nb\rangle} +\mathcal{P}(2,3,\ldots,n)
\end{equation}
where $K_{1,m} = \sum_{i=1}^m k_i$, and $\mathcal{P}(2, \dots, n)$ represents all permutations of the first term on the indicated legs. Although not manifest, this $h$ is symmetric under $a\leftrightarrow b$ and under any permutation of the remaining legs $\{1,2,\ldots,n\}$. At four points the amplitude  reduces to 
\begin{equation}
M_4^{1-\rm loop}(1^+,2^+,3^+,4^+) = -\frac{i}{(4\pi)^2\cdot 120}\left(\frac{\kappa}{2}\right)^4\left(\frac{s_{12}s_{23}}{\langle 12\rangle\langle 23\rangle\langle 34\rangle\langle 41\rangle}\right)^2(s_{12}^2+s_{23}^2+s_{13}^2) 
\end{equation}
where $s_{ij} = \langle ij\rangle[ij]$.

\section{Collinear Limits}\label{sec:collinear}

In this section we use collinear splitting functions to show that the $w_{1+\infty}$ algebra is uncorrected at loop level in self-dual gravity. We take care to distinguish between holomorphic collinear limits, in which $z_{12} \to 0$ with $\zb_{12}$ fixed, and ``true'' collinear limits in which $z_{12}, \zb_{12} \to 0$ together. In either case one defines
\be P \equiv p_1 + p_2, \qquad t \equiv \frac{\omega_1}{\omega_1+\omega_2}. \ee
In the true collinear case we have $P \propto p_1 \propto p_2$, but in the holomorphic collinear case it is only the left spinors that are parallel,
\be \lam_P = \frac{\lam_1}{\sqrt{t}} = \frac{\lam_2}{\sqrt{1-t}}, \ \ \ \ \tilde\lam_P = \sqrt{t} \, \tilde\lam_1 + \sqrt{1-t} \, \tilde\lam_2. \ee
In any theory, as two outgoing massless particles become collinear (in either sense) in a tree-level momentum-space amplitude it diverges in a universal fashion, described by a splitting function: 
\be \label{eq:split} M_n^{\rm tree}(1^{h_1}, 2^{h_2}, 3^{h_3}, \dots, n^{h_n}) \stackrel{1 \parallel 2} {\longrightarrow} \sum_{h_P = \pm} {\rm Split}_{h_P}^{\rm tree}(t, 1^{h_1}, 2^{h_2}) \times M^{\rm tree}_{n-1}(P^{-h_P}, 3^{h_3}, \dots, n^{h_n}). \ee
Here $h_i=\pm$ is the helicity of the $i$th particle. From~\cite{Pate:2019lpp}, the tree-level splitting functions for Einstein gravity are
\be \label{eq:split} \begin{aligned} {\rm Split}_+^{\rm tree}(t, 1^+, 2^+) & = 0, \\
{\rm Split}_-^{\rm tree}(t, 1^+, 2^+) & =  \frac{-\kappa}{2t(1-t)} \frac{[12]}{\langle 12 \rangle} = \frac{-\kap}{2t(1-t)} \frac{\zb_{12}}{z_{12}}. \end{aligned} \ee
It was shown in~\cite{Bern:1998sv} that the exact amplitudes of self-dual gravity have the same true collinear splitting property as tree-level Einstein gravity, essentially because the three-point amplitudes of self-dual gravity are tree-exact. It turns out that one can retrace their steps in the holomorphic collinear limit and find the same result. Specifically, one can show that with $z_{12}$ small and $\zb_{12}$ fixed we have the following limits of the half-soft functions,
\be h(a, \{1, 2, 3, \dots, n\}, b) \to \frac{1}{t(1-t)} \frac{[12]}{\langle 12 \rangle} \, h(a, \{P, 3, \dots, n\}, b), \ee
\be h(1, \{2, 3, \dots, n\}, b) \to \frac{\sqrt{t} \, \langle{b} |K_{3,n} |2]}{(1-t) \langle 12 \rangle \langle Pb \rangle} h(1, \{3, \dots, n\}, b), \ee
and that these properties imply that the full amplitude obeys
\be M_n^{1-\rm loop}(1^+, 2^+, \dots, n^+) \to \frac{-\kap}{2t(1-t)} \frac{[12]}{\langle 12 \rangle} M_{n-1}^{1-\rm loop}(P^+, \dots, n^+). \ee
In the course of this derivation one finds an additional term in the amplitude proportional to $\frac{[12]^3}{\langle 12 \rangle}$. In the true collinear limit it is trivially zero because $[12]$ is small. It turns out to also vanish in the holomorphic collinear limit due to the following identity,
\be 0 = \sum_{\stackrel{3\le b\le n}{M, N}} \langle Pb \rangle^2 h(P, M, b) h(b, N, P) \langle b | K_M |P] \langle P | K_N |b]^3, \ee
which was shown for $n\leq 13$ in~\cite{He:2014bga} and proven for all $n$ in~\cite{Rao:2016tgx}.

Now that we have the splitting function, we can proceed to derive the celestial OPE as in~\cite{Pate:2019lpp}. Note that unstripped amplitudes, which we will denote by $\mathcal{M}_n$, obey the same splitting formula as the stripped amplitudes $M_n$, because $\delta^{(4)}(p_1+p_2+\dots) = \delta^{(4)}(P+\dots)$. Celestial amplitudes containing only massless particles are obtained by Mellin transforming the corresponding momentum-space amplitudes~\cite{Pasterski:2016qvg}. Accordingly the all-plus celestial amplitude is\footnote{Some such amplitudes have been calculated explicitly in~\cite{Albayrak:2020saa,Gonzalez:2020tpi}.}
\be \widetilde{\mathcal{M}}_n(z_1, \zb_1, \Delta_1; \ldots; z_n, \zb_n, \Delta_n) = \left[\prod_{i=1}^n\int_0^\infty d\omega_i \omega_i^{\Delta_i-1}\right] \mathcal{M}_n(1^+,\ldots,n^+). \ee
Changing integration variables from $\omega_1, \omega_2$ to $\omega_P = \omega_1 + \omega_2$ and $t = \omega_1/\omega_P$, taking $z_{12}$ small, and applying the splitting formula within the integral gives
\begin{equation}
\begin{aligned}
& \widetilde{\mathcal{M}}_n(z_1, \zb_1, \Delta_1;\ldots;z_n,\zb_n,\Delta_n) \stackrel{z_{12} \to 0}{\longrightarrow} -\frac{\kap\bar{z}_{12}}{2z_{12}}\int_0^1 dt \, t^{\Delta_1-2}(1-t)^{\Delta_2-2}\cr
&\times  \int_0^\infty d\omega_P \,\omega_P^{\Delta_1+\Delta_2-1} \left[\prod_{i=3}^n\int_0^\infty d\omega_i \omega_i^{\Delta_i-1}\right]  \mathcal{M}_{n-1}(P^+,3^+,\ldots, n^+) + \mathcal{O}(z_{12}^0) \cr
&= -\frac{\kap\bar{z}_{12}}{2z_{12}}\int_0^1 dt \, t^{\Delta_1-2}(1-t)^{\Delta_2-2} \widetilde{\mathcal{M}}_{n-1}(z_2, \zb_2 + t\zb_{12}, \Delta_1+\Delta_2; \dots; z_n, \zb_n, \Delta_n) + \mathcal{O}(z_{12}^0).
\end{aligned}
\end{equation}
Now we Taylor expand the $t$ dependence in $\widetilde{\mathcal{M}}_{n-1}$ and recognize the resulting $t$ integrals as Euler beta functions
\begin{equation}
\begin{aligned}
& \widetilde{\mathcal{M}}_n(z_1, \zb_1, \Delta_1;\ldots;z_n,\zb_n,\Delta_n) \stackrel{z_{12} \to 0}{\longrightarrow}  -\frac{\kap\bar{z}_{12}}{2z_{12}} \sum_{n=0}^\infty \left( \int_0^1 dt \, t^{\Delta_1-2+n}(1-t)^{\Delta_2-2} \right)\cr
 &\times  \frac{\zb_{12}^n}{n!} \bar\p^n \widetilde{\mathcal{M}}_{n-1}(z_2, \zb_2, \Delta_1+\Delta_2; \dots; z_n, \zb_n, \Delta_n) + \mathcal{O}(z_{12}^0) \cr
&=  -\frac{\kap\bar{z}_{12}}{2z_{12}} \sum_{n=0}^\infty B(\Delta_1-1+n,\Delta_2-1) \frac{\zb_{12}^n}{n!} \bar\p^n \widetilde{\mathcal{M}}_{n-1}(z_2, \zb_2, \Delta_1+\Delta_2; \dots; z_n, \zb_n, \Delta_n) + \mathcal{O}(z_{12}^0) .
\end{aligned}
\end{equation}
From this equation we can read off the singular part of the celestial OPE of $G^+_{\Delta_1} G^+_{\Delta_2}$. We have 
\begin{equation}
G_{\Delta_1}^+(z_1,\bar{z}_1)G_{\Delta_2}^+(z_2,\bar{z}_2)\sim -\frac{\kappa}{2z_{12}} \sum_{n=0}^\infty B(\Delta_1-1+n,\Delta_2-1)\frac{\bar{z}_{12}^{n+1}}{n!}\bar{\partial}^nG_{\Delta_1+\Delta_2}^+(z_2,\bar{z}_2) 
\end{equation}
which is precisely~\eqref{eq:GOPE}, but now this is an exact result in quantum self-dual gravity as opposed to a tree-level result. From here one can plug in $H^k(z,\zb) = \lim_{\varepsilon \to 0} \varepsilon G^+_{k+\varepsilon}(z,\zb)$ to obtain
\begin{equation}
\begin{aligned}
H^k(z_1,\bar{z}_1)&H^\ell(z_2,\bar{z}_2) \sim \cr 
&\lim_{\varepsilon_1,\varepsilon_2\rightarrow 0} -\frac{\kappa \varepsilon_1\varepsilon_2}{2z_{12}} \sum_{n=0}^\infty \frac{\Gamma(k+\varepsilon_1-1+n)\Gamma(\ell+\varepsilon_2-1)}{\Gamma(k+\ell+\varepsilon_1+\varepsilon_2-2+n)}\frac{\bar{z}_{12}^{n+1}}{n!}\bar{\partial}^n G^+_{k+\ell+\varepsilon_1+\varepsilon_2}(z_2,\bar{z}_2)\cr
& = -\frac{\kappa}{2z_{12}} \sum_{n=0}^{1-k} \binom{2-k-\ell-n}{1-\ell} \frac{\bar{z}_{12}^{n+1}}{n!} \bar{\partial}^n H^{k+\ell}(z_2,\bar{z}_2)
\end{aligned}
\end{equation}
which is~\eqref{eq:fullHOPE}. 

Contour integrating with respect to $\bar{z}_1,\bar{z}_2$ on both sides and using~\eqref{eq:Hcont} one recovers, after some algebra (for details see Appendix A of~\cite{Guevara:2021abz}), the OPE~\eqref{eq:HmOPE} of the $H^k_n(z)$-currents. As per~\cite{Strominger:2021lvk}, the mode redefinition~\eqref{wdef} then gives the current algebra~\eqref{eq:wOPE} 
\be \label{eq:wOPE2}w^p_m(z_1) w^q_n(z_2) \sim \frac{[m(q-1) - n(p-1)] w^{p+q-2}_{m+n}(z_2)}{z_{12}}, \ee
which immediately implies that the celestial commutator~\eqref{eq:wcom} is undeformed in the quantum theory.

\acknowledgments

We are grateful to Tim Adamo, Kevin Costello, Alfredo Guevara, Indranil Halder, Mina Himwich, Lionel Mason, Ricardo Monteiro, \mbox{Monica} Pate and Atul Sharma for useful discussions. The work of JS was supported by the F.R.S.-FNRS Belgium through the
convention IISN 4.4503 and by the Erwin-Schr\"{o}dinger fellowship J-4135 of the Austrian Science Fund (FWF) during the early stages of this work. This work was supported by DOE grant \mbox{de-sc/0007870}, and  the Gordon and Betty Moore \mbox{Foundation} and the John Templeton Foundation via the Black Hole Initiative.

\bibliography{walg}
\bibliographystyle{JHEP}

\end{document}